# Attractive double-layer forces between neutral hydrophobic and neutral hydrophilic surfaces

Eduardo R. A. Lima[1,2,*], Mathias Boström[3], Nadine Schwierz[4], Bo E. Sernelius[3,*], and Frederico W. Tavares[2,5]

[1]Instituto de Química, Universidade do Estado do Rio de Janeiro, PHLC, CEP 20550-900, [2]Programa de Engenharia Química, COPPE, Universidade Federal do Rio de Janeiro, 21945-970, Rio de Janeiro, RJ, Brazil, [3]Division of Theory and Modeling, Department of Physics, Chemistry and Biology, Linköping University, SE-581 83 Linköping, Sweden, [4]Physik Department, Technische Universität München, 85748 Garching, Germany, [5]Escola de Química, Universidade Federal do Rio de Janeiro, Cidade Universitária, CEP 21949-900, Rio de Janeiro, RJ, Brazil.

The interaction between surface patches of proteins with different surface properties has a vital role to play driving conformational changes of proteins in different salt solutions. We demonstrate the existence of ion-specific attractive double-layer forces between neutral hydrophobic and hydrophilic surfaces in the presence of certain salt solutions. This is done by solving a generalized Poisson-Boltzmann equation for two unequal surfaces. In the calculations we utilize parameterized ion-surface-potentials and dielectric-constant-profiles deduced from recent non-primitive-model molecular dynamics (MD) simulations that account partially for molecular structure and hydration effects.

* Corresponding authors: bos@ifm.liu.se (BES); lima.eduardo@gmail.com (ERAL)

# I. Introduction

In this work we explore a previously not recognized contribution to the attraction between an uncharged hydrophobic and an uncharged hydrophilic surface. Fig. 1 presents a schematic illustration of such hydrophobic and hydrophilic surfaces interacting across salt solution. Ion specific double layers of opposite signs may be set up at the two surfaces leading to an attractive force. We propose that this may be important to account for when considering the interaction between surface patches of different proteins having different surface properties [1]. Such an effect was predicted by Schwierz et al. based on their calculated double layers that had opposite signs at a single hydrophobic surface as compared to a single hydrophilic surface [2]. Our approach is to solve a generalized Poisson-Boltzmann equation [3] for two unequal planar surfaces. In this solution water is treated as a continuum. However, we partially account for molecular structure [2,4], as well as dispersion potentials [5] and hydration effects [6] by utilizing ion-surface parameterized potentials of mean force (PMF) and dielectric constant profiles deduced from very recent non-primitive model MD simulations [2].

We show that ion specific double layer force is similar in magnitude to the attractive Hamaker-van der Waals force at moderate to high salt concentrations (around 0.1-1 M). It clearly gives an important contribution to the short-range attraction. We note that previous work on specific ion effects using PMF from simulations have dealt with a single surface or with the interaction between two equal surfaces. This is the first paper that addresses the ion specific double layer interaction between a hydrophobic and a hydrophilic surface in the presence of different salt solutions.

The structure of this article is as follows. In Section II, we briefly present the basic theory of the generalized Poisson-Boltzmann equation and include a very short recapitulation of the parameterized PMFs used. We also give the asymptotic expression for the Hamaker-van der Waals force between two self-assembled monolayers (SAMs) that will be used as comparison with the double layer pressure; the expression is appropriate when the surfaces are not so close that they overlap but at the same time not so far apart that the finite thickness of the SAM becomes relevant. Then, in Section III we present the results of our calculations for two unequal surfaces. These results demonstrate that attractive double layer forces are set up between an uncharged hydrophobic and an uncharged hydrophilic surface in the presence of certain salt solutions. Finally we end with some concluding remarks in Section IV.

**II. Theory**

In the classical DLVO theory van der Waals-Lifshitz forces are treated separately, in a linear inconsistent way [5]. In such standard theory, the Poisson-Boltzmann (PB) equation takes into account the electrostatic potential, only. After the solution is obtained, the direct van der Waals interaction between the two planar surfaces is added. This ansatz is incomplete and the theory cannot predict ion specificity, which is commonly observed experimentally [5]. In a more complete theory, additional non-electrostatic (NES) interactions [5-8] and the effect of short-range hydration [6] must be treated at the same level as the electrostatic forces acting on ions.

Extending earlier work that considered two equal surfaces [9], we use an ion specific generalized Poisson-Boltzmann equation for two unequal, neutral surfaces a distance $L$ apart:

$$\varepsilon_0 \frac{d}{dx}\left(\varepsilon(x)\frac{d\psi}{dx}\right) = -e\sum_i c_{0,i} z_i \exp\left[-(z_i e\psi + U_i(x))/k_B T\right], \quad (1)$$

$$\left.\frac{d\psi}{dx}\right|_{x=0} = 0, \quad \left.\frac{d\psi}{dx}\right|_{x=L} = 0, \quad (2)$$

where $k_B$ is the Boltzmann constant, $z_i$ is the valency of the ions, $e$ is the elementary charge, $T$ is the temperature of the system, $\psi$ is the self consistent electrostatic potential, $U_i(x) = W^a(x) + W^b(L-x)$ is the sum of the ionic potentials (PMF) acting between the ion and each of the two surfaces (surface '$a$' is hydrophobic and surface '$b$' is hydrophilic). Schwierz et al. performed simulations that gave the PMF for ions, and $\varepsilon(x)$ shown in Figures 2 and 3, near a single surface [2]. The local dielectric constant $\varepsilon(x)$ is defined as a function of the position $x$ from the hydrophobic surface. Following our previous work we define the local dielectric constant $\varepsilon(x)$ as a function of the position $x$ based on density simulation results, considering that these two quantities are highly correlated [2,9]. For the interaction between two surfaces, an important issue is how to treat the overlap of the dielectric functions of both surfaces for short distances between the plates. Because of that we do not consider any calculation for separating distances ($L$) shorter than 1.5 nm. For distances between 1.5 and 3 nm (where there is overlap), we choose carefully an adequate point of intersection in order not to take away important parts of these curves. In order to clarify this question we show in Figure 2 the local dielectric constant as a function of the position ($x$) between the surfaces for $L = 1.5$ nm where there is overlap of the original dielectric functions of single plates. In this case we chose $x = 1$ nm as the connection point. One can notice that in this point there is a very small discontinuity that does not affect the numerical results presented in other figures. We base our study on realistic NES potentials based on parameterized PMFs and dielectric profiles obtained by molecular dynamics simulations of ions in solution near an uncharged hydrophobic or an uncharged hydrophilic surface [2]. PMFs for three different ions between these two surfaces are reproduced in Fig. 3 [2]. We observe that the cations ($Na^+$) are attracted towards the hydrophilic surface while anions (at least $I^-$) are attracted towards the hydrophobic surface. This leads to opposite signs of ion specific double layers formed between the two different surfaces.

The electrostatic potential profiles and ion distributions obtained from the solution of Eq. (1) are used to calculate the pressure between the surfaces. The double layer pressure between two planar surfaces at a distance $L$ can be calculated by the differentiation of the free energy of the system [9]:

$$P = -\frac{\partial}{\partial L}\left(\frac{A}{area}\right), \tag{3}$$

and the free energy per unit of area is expressed by [9]

$$\frac{A}{area} = \frac{e}{2}\int_0^L \psi \sum_i c_i z_i \, dx + \int_0^L \sum_i c_i U_i \, dx + k_B T \int_0^L \sum_i c_i \left[\ln\left(\frac{c_i}{c_{0,i}}\right) - 1\right] dx. \tag{4}$$

The first two terms on the right hand side of Eq. (4) are the energy contributions (electrostatic and the ionic potential of mean force contribution, respectively) to the free energy of the system and the third term is the entropic contribution.

At large surface to surface distances where the ion-surface interaction potential approaches zero, Eq. (1) can be linearized corresponding to the Debye-Hückel theory. The solution of the linearized equation is

$$\psi(x) = \psi_{DH} \exp(-\kappa x), \qquad (5)$$

where $\kappa$ is the inverse Debye screening length, $\kappa = \sqrt{\left(e^2 \sum_i c_{0,i} z_i^2\right) / \varepsilon \varepsilon_0 k_B T}$. The resulting Debye-Hückel surface potential $\psi_{DH}$ is a measure of the effective surface charge density $\sigma_{DH}$,

$$\sigma_{DH} = \varepsilon \varepsilon_0 \kappa \psi_{DH}. \qquad (6)$$

The term "Debye-Hückel surface potential" is a little misleading, since this may lead the reader to think that the potential $\psi_{DH}$ is the potential that one would find from the Debye-Hückel theory of the double layer. In fact, the potential $\psi_{DH}$ results from ion-specific effects at the hydrophilic/hydrophobic surface [2], and not from the standard Debye-Hückel theory. All ion-specific effects are included in the term $\psi_{DH}$ and are assumed to be relevant only right at the surface. At all other distances, the "far-field" Debye-Hückel solution for the potential [Eq. (5)] is used.

The effective surface charge densities allow one to predict the long-range forces between two surfaces. At large surface separations where the exponentially decaying electrostatic potential is small, the pressure between the two surfaces can be calculated from [10]

$$P(L) = \frac{\left(2\sigma_{DH}^1 \sigma_{DH}^2\right)\left(2 + e^{L\kappa} + e^{-L\kappa}\right)}{\varepsilon \varepsilon_0 \left(e^{L\kappa} - e^{-L\kappa}\right)^2}, \qquad (7)$$

where $\sigma_{DH}^j$ is the effective surface charge density of the two surfaces, which is calculated from the Debye-Hückel surface potential [2]. Here we use the bare dielectric constant of bulk water since the point of interest is the midplane between two remote surfaces.

The van der Waals-Hamaker pressure between two planar surfaces coated with SAMs is (at short range) approximately given by [11]

$$P(L) = -\frac{A_{12}}{6\pi L^3}, \tag{8}$$

where $A_{12}$ is the Hamaker constant between two self-assembled monolayers. Here we use $A_{12} = 1.21\ k_B T$ based on the value reported by Kokkoli and Zukoski [12] for two SAMs of hexadecanethiol.

We calculate the Hamaker contribution to the pressure just to compare its magnitude to the double layer pressure obtained by the modified Poisson-Boltzmann theory. The van der Waals force is always present and plays an essential role in all phenomena involving intermolecular forces. Hence, the van der Waals force is a good reference function that can be used to check whether the magnitude of any other contribution is significant or not.

### III. Results

In Fig. 4a we explore the difference between the concentrations of cations and anions for 0.1 M NaI and 0.1 M NaCl. This quantity is proportional to the charge density profile between the plates. Ion specific double layers of opposite signs are set up at the two surfaces in the presence of NaI. We show the corresponding curves for different concentrations of NaI in Fig. 4b. This would suggest the possibility of an attractive pressure between these surfaces that increases with increasing salt concentration. The reader is referred to the cited paper (and the supporting materials of that paper) for simulation details [2].

In Fig. 5 we consider how the pressure between an uncharged hydrophilic and an uncharged hydrophobic surface depends on the specific choice of salt ions in the 1 M salt solution used (NaCl (dash-dotted line) and NaI (solid line)). At moderate to high salt concentrations the effect is similar in magnitude to the attractive van der Waals-Hamaker pressure between the two surfaces (dashed line). We observe that the attraction between the two asymmetric surfaces increases with increasing size (and polarizability) of the anions. The double layer contribution is important and highly ion specific with stronger attraction for NaI (around 4 times the Hamaker attraction between the surfaces) and a weaker attraction for NaCl (smaller than the Hamaker interaction).

In Fig. 6 we show the comparison between the pressure calculated from PB theory and the pressure calculated from the effective surface charge densities using Eq. (7) as a function of the separation between the plates for 1M salt concentration. The agreement is perfect for large distances even for this high concentration. At the two different interfaces ion specific adsorption leads to effectively charged surfaces, which interact via screened Coulomb interactions with each other. In NaI solutions the long-range potential and therefore the effective surface charge density is negative at the hydrophobic surface but positive at the hydrophilic surface [2]. This results in a long-range attraction and therefore a destabilization of uncharged pairs of hydrophobic/hydrophilic surfaces in NaI solutions. For NaCl solutions a long-range attraction is predicted as well, which is much smaller than for NaI due to the smaller effective surface charges on both surfaces.

The question raised by our work is if the double layer pressure between different combinations of hydrophobic and hydrophilic SAMs are comparable in magnitude to the van der Waals-Hamaker pressure for 0.1 M salt solutions. Since the pressures are higher for NaI we only present the result for this salt solution. We see in Fig. 7 that the double layer pressure at 0.1 M salt concentration is smaller in magnitude as compared to the double layer pressure at 1 M salt concentration (Fig. 5). This is consistent with Fig. 4 that demonstrates that the double layers get more pronounced with increasing salt concentration. There is also an increasing screening effect with increasing salt concentrations but this seems to be less important. We see that the repulsive double layer pressure between two hydrophobic surfaces is similar in magnitude to the van der Waals-Hamaker pressure regardless of plate separation. However, the pressure between one hydrophobic surface and one hydrophilic surface is weakly repulsive at shorter distances and then weakly attractive at larger. The double layer repulsion between two hydrophilic surfaces is similar in magnitude to the Hamaker force for distances larger than 2 nm.

In Fig. 8 we consider a 0.5 M salt solution of NaI. We see that for this moderate salt concentration, there is strong double layer repulsion between two hydrophobic surfaces due to large adsorption of iodide on both surfaces. For two hydrophilic surfaces there is weak long-range repulsion and a weak short-range attraction. More importantly, at this concentration we observe an attraction between one hydrophobic and one hydrophilic surface, which is of the same order of magnitude as the Hamaker attraction regardless of the separation distance.

**IV    Conclusions**

The electrostatics of charged particle adsorption (e.g. protein) to surfaces [13,14] and interaction of dissimilar charged surfaces [15] have been explored in the past. However, our paper addresses for the first time the ion specific double layer attraction between an uncharged hydrophobic surface and an uncharged hydrophilic surface. Due to opposite signs of the double layers the ion specific double layer pressure between an uncharged hydrophilic surface and an uncharged hydrophobic surface can give a strong attraction that is similar in magnitude to the Hamaker-van der Waals pressure. It is of vital importance to take this effect into account at moderate to medium salt concentrations (0.1 - 1 M) where on the one hand the charge due to specific ion adsorption is not too small and on the other hand the screening length is not too short.

We have found an ion specific double layer attraction between unequal surfaces that has not been explored in the past. The attraction between the two asymmetric surfaces increases with increasing size (and polarizability) of the ions. This attraction is bound to be important for the interaction between patchy proteins with hydrophobic and hydrophilic surface patches [1].

**Acknowledgements**

ERAL and FWT acknowledge financial support from the FAPERJ, CAPES and CNPq (Brazilian Agencies). MB and BES acknowledge financial support from the VR-contract No:70529001. For valuable discussions, we thank Prof. Barry Ninham, Prof. Roland Netz, Dr. Drew Parsons, and Dr. Dominik Horinek.

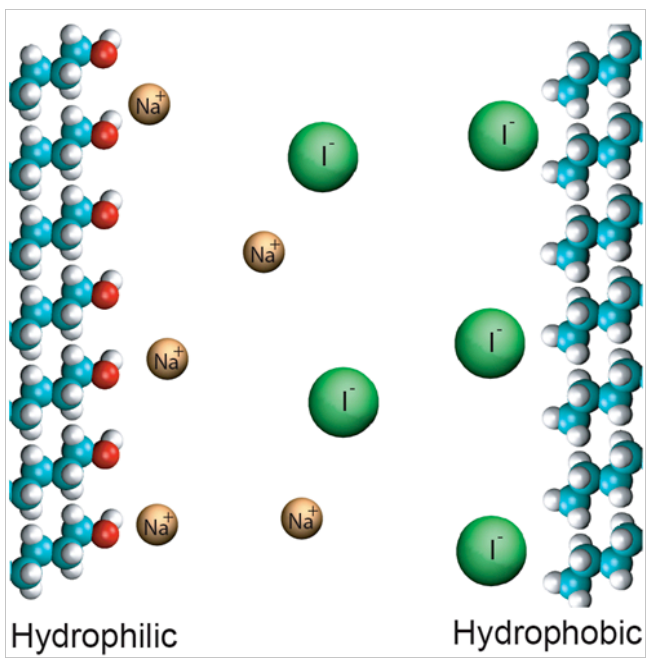

Fig. 1. (Color online) A schematic illustration of hydrophobic and hydrophilic surfaces interacting across salt solution. Opposite signs of the double layers can give rise to ion specific attraction.

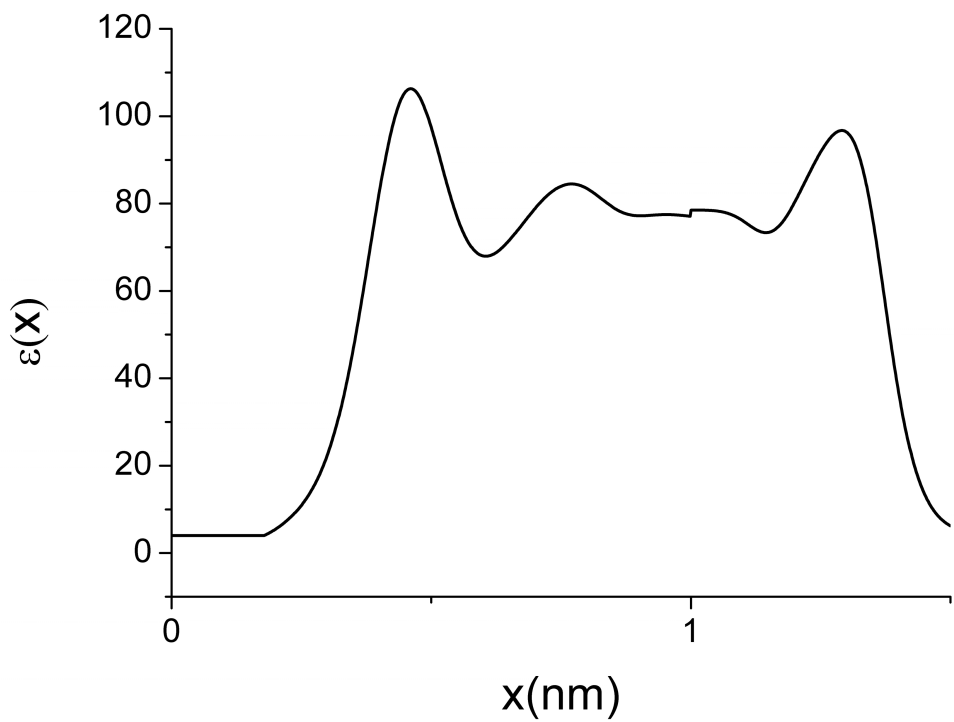

Fig. 2. Local dielectric constant as a function of the position for a surface-to-surface distance $L = 1.5$ nm. The hydrophobic surface is located at $x = 0$ and the hydrophilic surface at $x = 1.5$ nm.

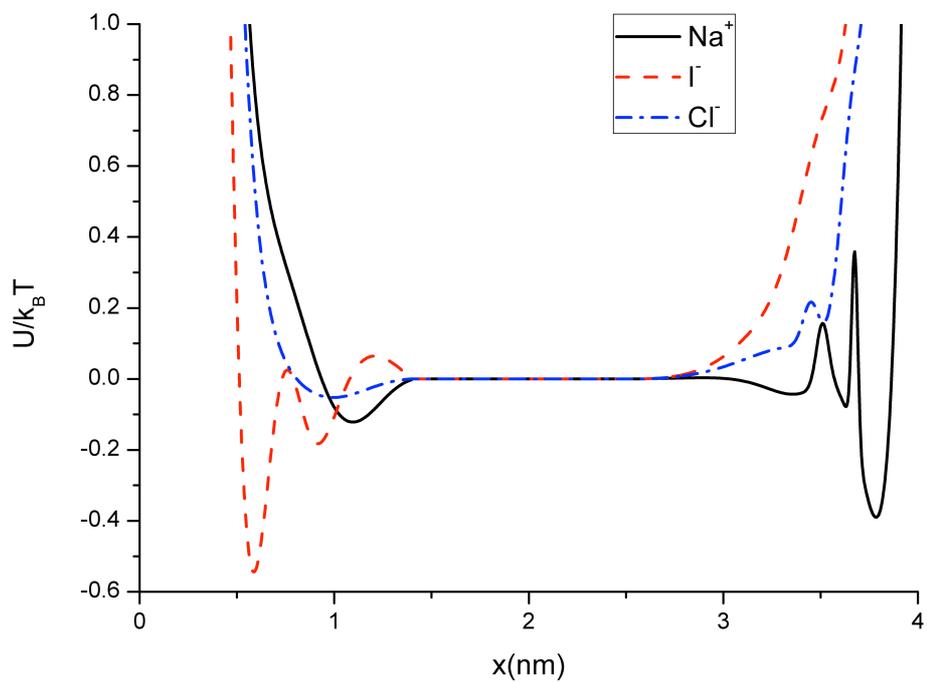

Fig. 3. (Color online) Parameterized PMFs between a hydrophobic surface at $x = 0$ nm and hydrophilic surface at $x = 4$ nm for different ions (using in all examples parameters given in the supporting material of ref. [2]).

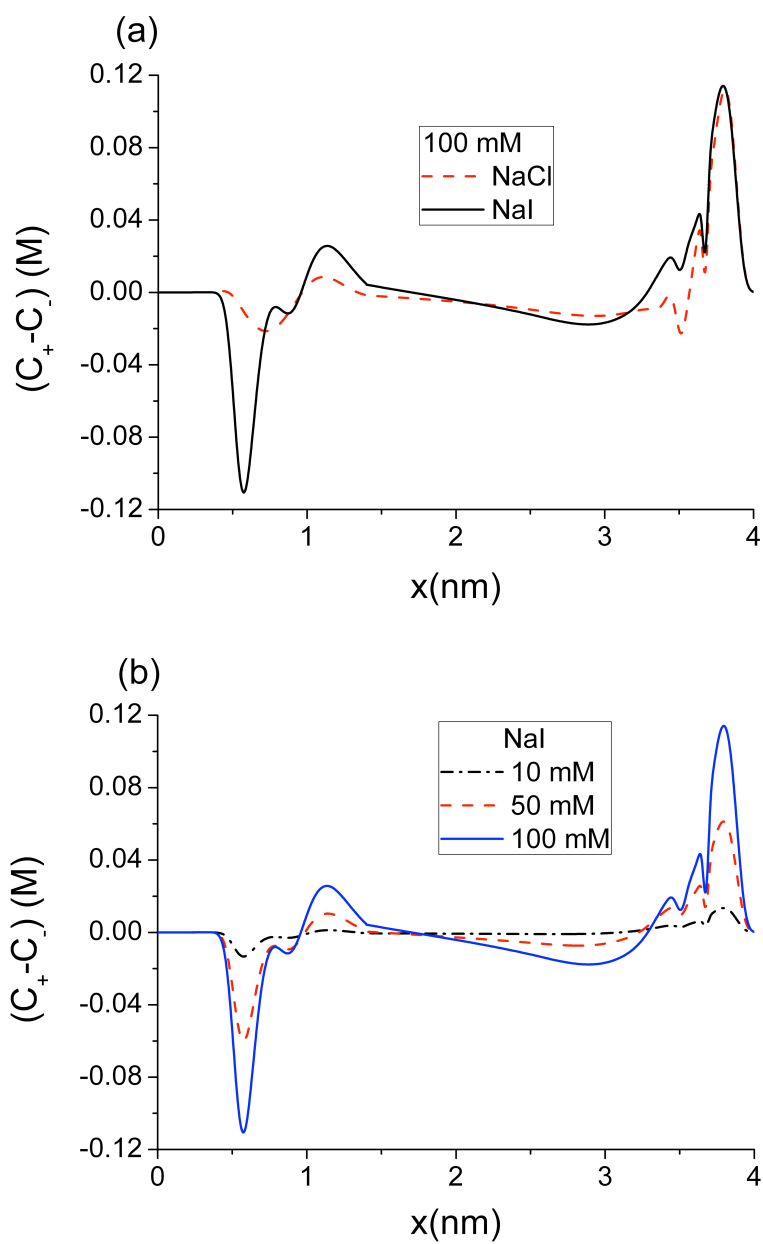

Fig. 4. (Color online) Difference between the concentration of cations ($C_+$) and anions ($C_-$) (proportional to the charge density profile) (a) for two different salt solutions: 0.1M NaCl (dashed line) and 0.1M NaI (solid line); (b) NaI at different concentrations: 0.01 M (dash-dotted line), 0.05 M (dashed line) and 0.1 M (solid line) between an uncharged hydrophobic surface (left) and an uncharged hydrophilic surface (right) 4 nm apart.

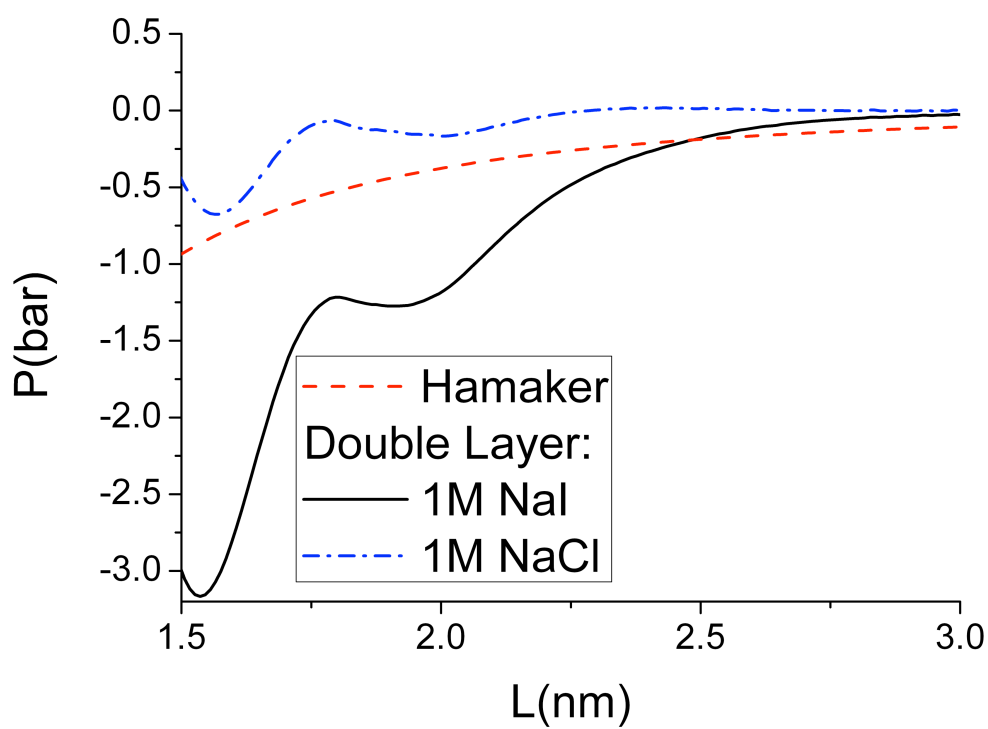

Fig. 5. (Color online) Van der Waals-Hamaker (dashed line) and double layer pressure as functions of separation, *L*, between an uncharged hydrophilic and a hydrophobic surface for 1M NaI (full line) and 1M NaCl (dash-dotted line) salt solutions.

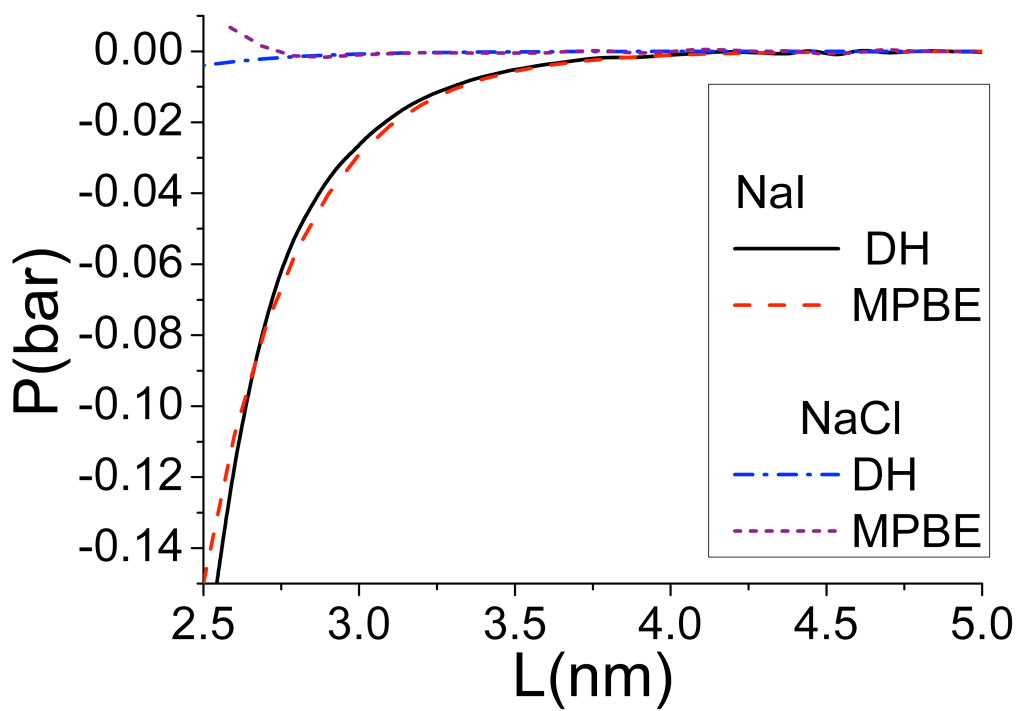

Fig. 6. (Color online) Comparison between the double layer pressure calculated from Eq. (7) in the Debye-Hückel limit and the corresponding results obtained from the numerical solution of the modified Poisson-Boltzmann equation for both salts at 1M. $L$ is the separation between the surfaces.

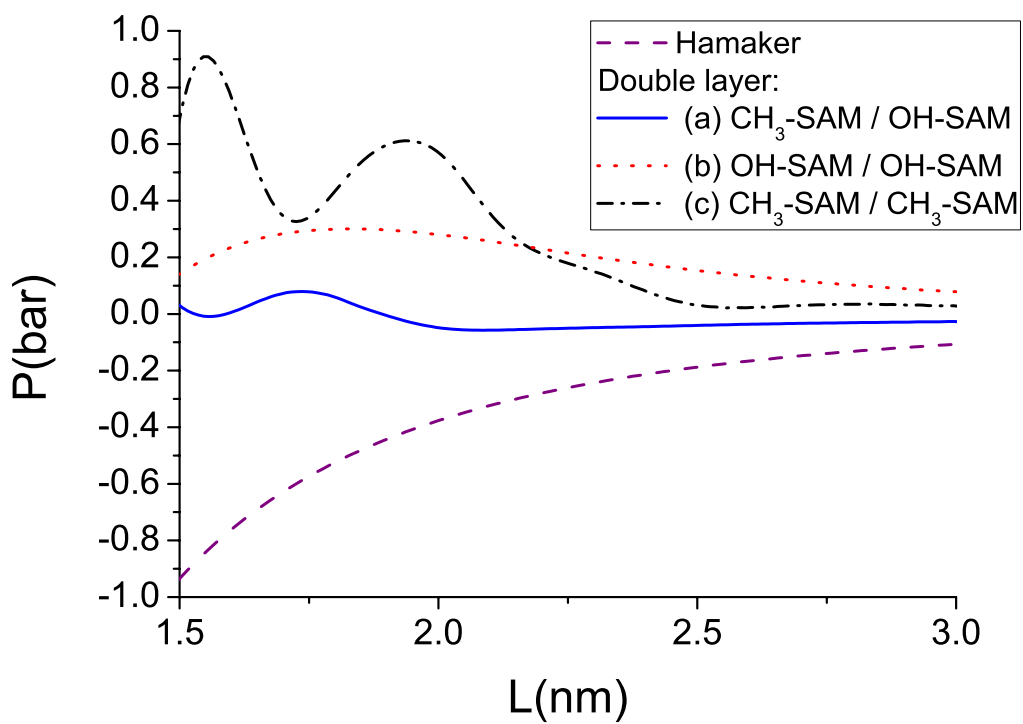

Fig. 7. (Color online) Van der Waals-Hamaker (dashed line) and double layer pressure in 0.1 M NaI between (a) an uncharged hydrophilic and a hydrophobic surface (full line), (b) two uncharged hydrophilic surfaces (dotted line), (c) two hydrophobic surfaces (dash-dotted line). $L$ is the separation between the surfaces.

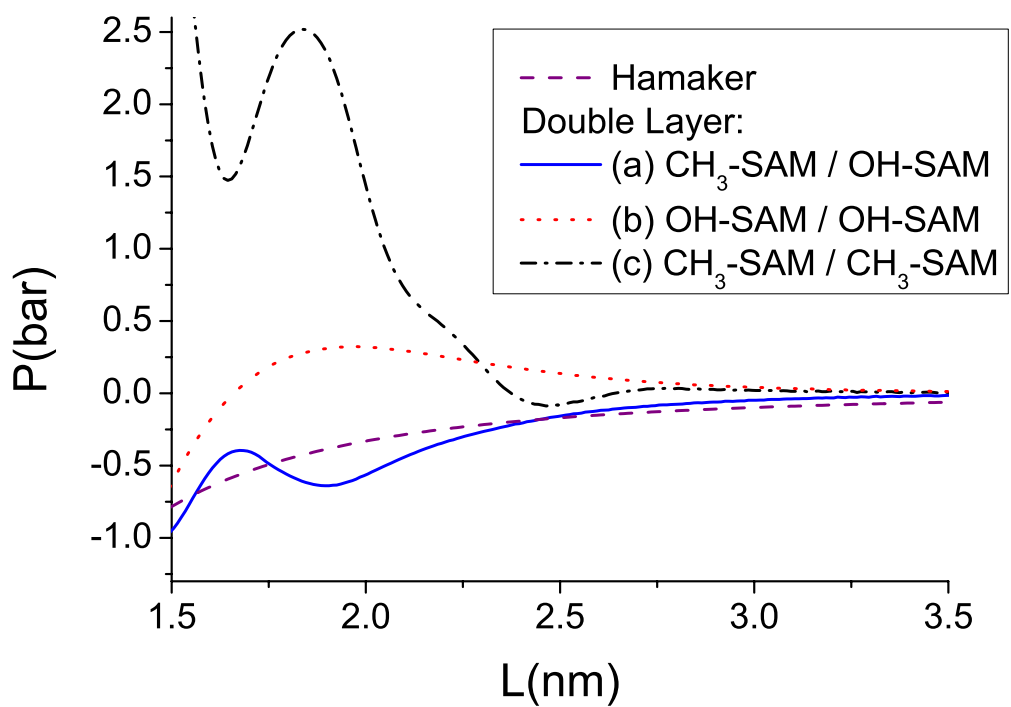

Fig. 8. (Color online) Same as Fig. 7, but for 0.5 M salt solution.